\documentclass[11pt,a4paper]{article}
\pdfoutput=1

\usepackage{jcappub}
\usepackage[utf8]{inputenc}
\usepackage{graphicx}
\usepackage{epsfig}
\usepackage{subfigure}
\usepackage{color}
\usepackage{comment}

\def\beq{\begin{equation}}
\def\eeq{\end{equation}}
\def\baq{\begin{eqnarray}}
\def\eaq{\end{eqnarray}}

\newcommand{\be}{\begin{equation}} 
\newcommand{\ee}{\end{equation}}
\newcommand{\bea}{\begin{equation}\begin{aligned}} 
\newcommand{\eea}{\end{aligned}\end{equation}}
\newcommand{\nn}{\nonumber}
\newcommand{\bmp}{\noindent\begin{minipage}{16cm}}
\newcommand{\emp}{\end{minipage}\vskip 7mm} 
\def\lsim{\mathrel{\raise.3ex\hbox{$<$\kern-.75em\lower1ex\hbox{$\sim$}}}}
\def\gsim{\mathrel{\raise.3ex\hbox{$>$\kern-.75em\lower1ex\hbox{$\sim$}}}}

\newcommand{\intron}[1]{}

\title{Resurrecting Quadratic Inflation with a non-minimal coupling to gravity}

\author[]{Tommi Tenkanen}

\affiliation[]{Astronomy Unit, Queen Mary University of London,
 \\ Mile End Road, London, E1 4NS, United Kingdom}
    
\emailAdd{t.tenkanen@qmul.ac.uk}

\abstract{We study Quadratic Inflation with the inflaton field $\phi$ coupled non-minimally to the curvature scalar $R$, so that the potential during inflation is of the form $V\propto m^2\phi^2+\xi R\phi^2$. We show that with a suitable choice of the non-minimal coupling strength, $\xi=\mathcal{O}(10^{-3})$, one can resurrect the success of the scenario when compared against the Planck and BICEP2/Keck Array data, and that in the region of the parameter space which is still allowed the model predicts values of the tensor-to-scalar ratio in the range $0.01\leq r < 0.12$, making it possible to either confirm the scenario or rule it out already by the current or near-future experiments, such as BICEP3 or LiteBIRD. However, we show that in this case the near-future observations are unlikely to be able to distinguish between the metric and Palatini formulations of gravity.}

\keywords{Quadratic inflation, non-minimal coupling, Planck results, models of inflation, Palatini gravity}

\begin{document}
\maketitle


\section{Introduction}

Among the plethora of models of cosmic inflation, the chaotic inflation models with dynamics characterized by the potential of the form $V\propto \phi^n$ are perhaps the simplest ones \cite{Linde:1983gd,Martin:2013tda,Martin:2013nzq}. The Quadratic Inflation model, $V\propto \phi^2$, is particularly simple, containing only one model parameter, the mass of the inflaton field, $m$, which can be fixed by requiring the model to yield the measured amplitude for the primordial curvature power spectrum. Calculating predictions also for other observables, such as spectral index and tensor-to-scalar ratio, is relatively easy in this model, but the recent analysis of the Planck and BICEP2/Keck Array data shows that the model generally predicts a too large tensor-to-scalar ratio, $r\simeq 0.13$, the current $2\sigma$ upper bound being at $r< 0.12$ \cite{Ade:2015tva}.

After the first Planck results, different ways to rescue the model have been studied in the literature, for instance in Refs. \cite{Ellis:2013iea,Vennin:2015vfa,Chakravarty:2015yho}. Typically these models contain additional fields which might act as curvaton-like fields or supplementary inflatons. However, in this work we take a different approach and consider a scenario where the inflaton field is coupled non-minimally to gravity, $V\propto \xi R\phi^2$, where $R$ is the curvature scalar. 

Applying this possibility to inflation is not a new idea, as non-minimal couplings to gravity have been discussed in a large number of works over the past decades, for instance in Refs. \cite{Futamase:1987ua,Salopek:1988qh,Fakir:1990eg,Amendola:1990nn,Kaiser:1994vs,Bezrukov:2007ep,Bauer:2008zj,Park:2008hz,Linde:2011nh,Kallosh:2013maa,Kallosh:2013tua,Chiba:2014sva,Boubekeur:2015xza,Pieroni:2015cma,Salvio:2017xul}. Despite the fact that usually inflation is studied merely as an effective field theory, in concrete model frameworks such a non-minimal coupling to gravity should not be seen as an {\it ad hoc} addition to the model but as a natural ingredient generated by quantum corrections in a curved space-time \cite{Birrell:1982ix}. In particular, this is the case for the scenario where the Higgs field of the Standard Model of particle physics (SM) acts as the inflaton field, a scenario made famous by Ref. \cite{Bezrukov:2007ep}.

In this paper, we do not consider any concrete model frameworks but adopt the effective theory point of view -- with an interesting generalization in the gravity sector: instead of assuming the usual metric case, where the connection is the Levi-Civita one, we also allow for the connection to be an independent variable, i.e. study the dynamics also in the context of {\it Palatini gravity}. Even though the metric and Palatini formalisms coincide within the theory of General Relativity, in more general models, especially in the ones where matter fields are coupled non-minimally to gravity, these two formalisms lead to two inherently different theories of gravity \cite{Sotiriou:2008rp}. In particular, this means that models of inflation with non-minimal couplings to gravity cannot be characterized just by the form of the matter field potential, but one needs to specify also the fundamental gravitational degrees of freedom, as was originally pointed out in Ref. \cite{Bauer:2008zj}.

Our work therefore contains three novel aspects: a) we study for the first time the Quadratic Inflation model with a non-minimal coupling to gravity in the context of not only metric but also Palatini gravity, b) we compare these models to the Planck and BICEP2/Keck Array data and show that with a suitable choice of the non-minimal coupling strength one can still resurrect the success of the scenario, and c) we show that in the region of the parameter space where the scenario is in agreement with the Planck data, the near-future experiments are unlikely to be able to distinguish between the metric and Palatini theories of gravity. 

Interestingly, we show that only a small non-minimal coupling to gravity, $\xi=\mathcal{O}(10^{-3})$, suffices to satisfy the Planck and BICEP2/Keck Array bounds. This is in contrast to e.g. the usual case of the SM Higgs inflation, where the non-minimal coupling is typically required to be very large, $\xi\simeq 10^4$, in the metric case and $\xi\simeq 10^9$ in the Palatini case (for recent studies on similar scenarios with a small non-minimal gravity coupling, see e.g. \cite{Tenkanen:2016twd,Alanne:2016mpa}). We also show that it is possible to test the scenario already with some currently on-going experiments, and thus either confirm it or rule it out in the near future.

The paper is organized as follows: in Sec. \ref{quadr_inf_revisited} we revisit the standard minimally-coupled Quadratic Inflation model, in Sec. \ref{quadr_inf_nonmin} we study the non-minimally coupled case, and then present our results in Sec. \ref{results}. Finally, we conclude in Sec. \ref{conclusions}.

\section{Quadratic Inflation revisited}
\label{quadr_inf_revisited}

In the standard Quadratic Inflation model the potential is given by 
\be
V = \frac{1}{2}m^2\phi^2 ,
\ee
where $\phi$ is the inflaton field and $m$ a mass term. Assuming slow-roll, $\dot{\phi}^2\ll V, |\ddot{\phi}|\ll 3H|\dot{\phi}|$, where $H$ is the Hubble scale and dots denote derivatives with respect to cosmic time, inflationary  dynamics is characterized by the usual slow-roll parameters
\begin{equation}
\label{SRparameters}
\epsilon \equiv \frac{1}{2}M_{\rm P}^2 \left(\frac{1}{V}\frac{{\rm d}V}{{\rm d}\phi}\right)^2 \,, \quad
\eta \equiv M_{\rm P}^2 \frac{1}{V}\frac{{\rm d}^2V}{{\rm d}\phi^2} \,,
\end{equation}	
where $M_{\rm P}$ is the reduced Planck mass, and the number of e-folds
\begin{equation}
N = \frac{1}{M_{\rm P}^2} \int_{\phi_f}^{\phi_i} {\rm d}\phi \, V \left(\frac{{\rm d}V}{{\rm d} \phi}\right)^{-1},
\label{Ndef}
\end{equation}	
where the field value at the end of inflation, $\phi_f$, is defined via $\epsilon(\phi_f) = 1$. For a given value of $N$, Eq. (\ref{Ndef}) determines the field value $\phi_i$ at the time the corresponding scales left the horizon. 

The correct amplitude for the curvature power spectrum, $\mathcal{P}_{\mathcal{R}}=(2.141\pm 0.052)\times 10^{-9}$ ($68\%$ confidence level) \cite{Ade:2015xua}, is obtained for \cite{Lyth:1998xn}
\begin{equation}
\label{cobe}
\frac{V(\phi_i)}{\epsilon(\phi_i)} = (0.027M_{\rm P})^4 ,
\end{equation}
from which we can fix the required mass to be $m\simeq 6\times 10^{-6}M_{\rm P}$ for $N=60$ and $m\simeq 7\times 10^{-6}M_{\rm P}$ for $N=50$. For the spectral index, $n_s$, and tensor-to-scalar ratio, $r$, we obtain
\begin{eqnarray}
\label{nsr}
n_s(\phi_i) &\simeq& 1+2\eta-6\epsilon \simeq 0.960 \dots 0.967 \\ \nn
r(\phi_i) &\simeq& 16\epsilon \simeq 0.13\dots 0.16
\end{eqnarray}
where the numerical values again apply for $N=50\dots 60$. As is well known, the results are slightly disfavored by the Planck data, which give $n_s=0.9681\pm 0.0044$ ($68\%$ confidence level) \cite{Ade:2015xua}, and by the joint analysis of Planck and BICEP2/Keck Array data, which gives $r<0.12$ ($95\%$ confidence level) \cite{Ade:2015tva}. However, in the next section we introduce a scenario which can effectively resurrect the Quadratic Inflation model.

\section{Quadratic Inflation with a non-minimal coupling to gravity}
\label{quadr_inf_nonmin}

We consider a theory which contains not only the usual kinetic and potential terms but which is non-minimally coupled to gravity
\be
\label{nonminimal_action}
S_J = \int d^4x \sqrt{-g}\left(\frac{1}{2} g^{\mu\nu}\partial_{\mu}\phi\partial_{\nu}\phi -\frac{1}{2}\left(M_{\rm P}^2 + \xi\phi^2\right) g^{\mu\nu}R_{\mu\nu}(\Gamma) - V(\phi)\right) ,
\ee
where $R_{\mu\nu}$ is the Ricci tensor, $\xi$ is a dimensionless coupling constant, $g_{\mu\nu}$ is the metric tensor and $g$ its determinant, and $\Gamma$ is the connection. In the metric formulation the connection is determined uniquely as a function of the metric tensor, i.e. it is the Levi-Civita connection $\bar{\Gamma}=\bar{\Gamma}(g^{\mu\nu})$, whereas in the Palatini formalism both $g_{\mu\nu}$ and $\Gamma$ are treated as independent variables, and the only assumption is that the connection is torsion-free, $\Gamma^\lambda_{\alpha\beta}=\Gamma^\lambda_{\beta\alpha}$.
	
\begin{figure}
\begin{center}
\includegraphics[width=.6\textwidth]{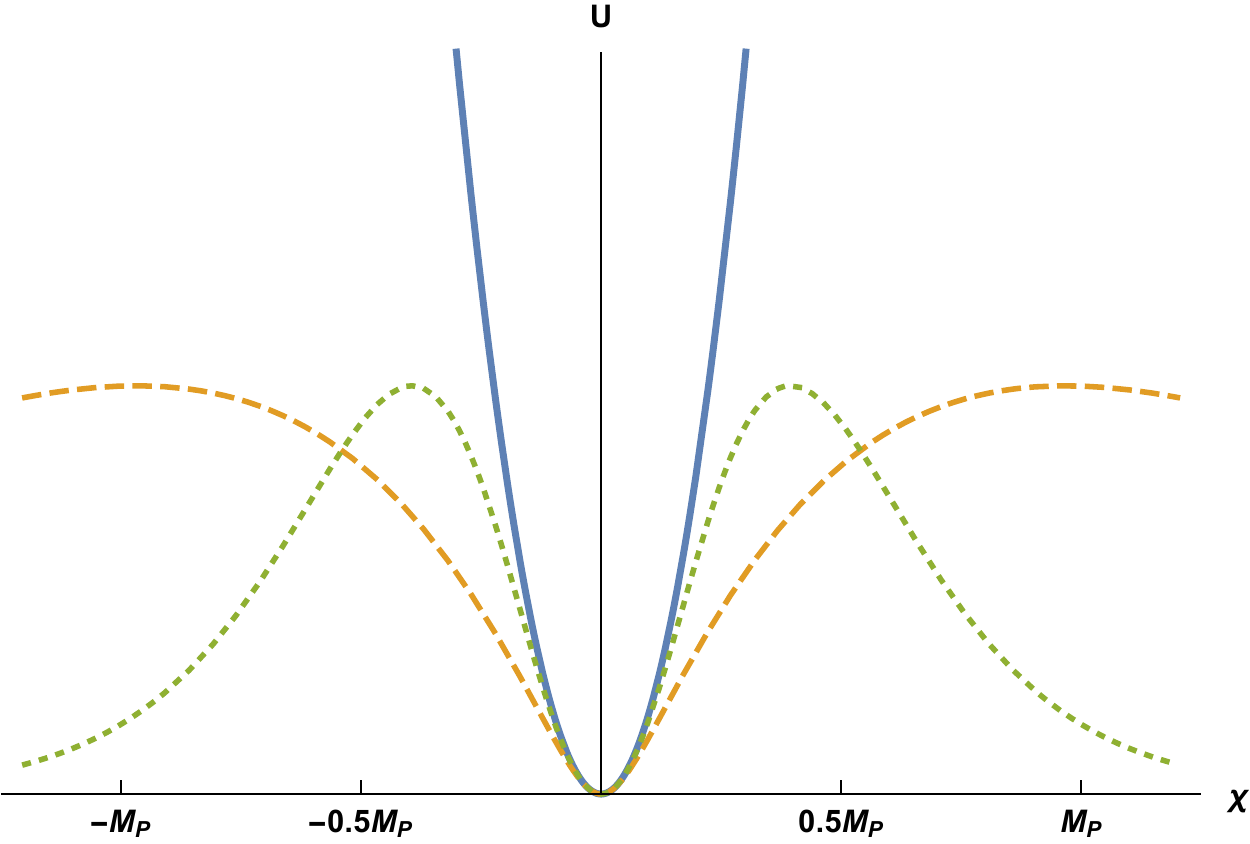}
\caption{An illustration of the inflaton potential in three different cases: Quadratic Inflation a) without a non-minimal coupling to gravity (solid blue line), b) with a non-minimal coupling to gravity in the metric case (dashed yellow line), and c) with a non-minimal coupling to gravity in the Palatini case (dotted green line). In this figure $\xi=5$.}
\label{potentials}
\end{center}
\end{figure}

The non-minimal coupling in the Jordan frame action \eqref{nonminimal_action} can be removed by a conformal transformation to the Einstein frame, 
\begin{equation}
\label{Omega}
g_{\mu\nu} \to \Omega(\phi)^{-1}g_{\mu\nu}, \hspace{.5cm} \Omega(\phi)\equiv 1+\frac{\xi \phi^2}{M_{\rm P}^2} \, ,
\end{equation}	
and the resulting expression can be brought into a canonically normalized form by redefining the field operator
\be
\label{chi}
\frac{d\phi}{d\chi} = \sqrt{\frac{\Omega(\phi)^2}{\Omega(\phi)+6f\xi^2\frac{\phi^2}{M_{\rm P}^2}}} \,,
\ee
where $f=1$ in the metric case and $f=0$ in the Palatini case. By these transformations, the action \eqref{nonminimal_action} becomes
\be
S_{\rm E} = \int d^4x \sqrt{-g}\bigg(-\frac{1}{2}M_{\rm P}^2R +\frac{1}{2}{\partial}_{\mu}\chi{\partial}^{\mu}\chi - U(\chi)  \bigg),
\label{EframeS}
\ee
where $U(\chi) = \Omega^{-2}(\phi(\chi))V(\phi(\chi))$ and $R = g^{\mu\nu}R_{\mu\nu}(\bar{\Gamma})$ independently of our initial choice of the gravitational degrees of freedom, i.e. whether the action \eqref{nonminimal_action} is written in the metric or Palatini formalism. This choice, however, affects the form of the potential, which in the Einstein frame becomes different in the two cases, as shown in Fig. \ref{potentials}. Indeed, the fact that the potentials are different in these three cases (minimally coupled $\phi^2$, non-minimal metric and Palatini cases) motivates us to study the observable consequences of such non-minimally coupled inflaton models.

Before discussing the results, let us study the Einstein frame potential $U(\chi)$ a little further. We start with the Palatini case, because in that case the definition \eqref{chi} allows to analytically solve for $\phi=\phi(\chi)$ \cite{Bauer:2008zj} 
\be
\label{chipalatini}
\frac{\sqrt{\xi}\chi}{M_{\rm P}} = {\rm sinh}^{-1}\left(\frac{\sqrt{\xi}\phi}{m_{\rm P}}\right) ,
\ee
where ${\rm sinh}^{-1}$ is the inverse hyperbolic sine function. As a result, the potential $U(\chi) = \Omega^{-2}(\phi(\chi))V(\phi(\chi))$ becomes
\be
\label{palatiniU}
U_{\rm P}(\chi) = \frac{m^2M_{\rm P}^2}{2\xi}\frac{{\rm sinh}^2\left(\frac{\sqrt{\xi}\chi}{M_{\rm P}}\right)}{\left(1+{\rm sinh}^2\left(\frac{\sqrt{\xi}\chi}{M_{\rm P}}\right)\right)^2} .
\ee
It is illustrative to write this as a series about the minimum, which gives
\be
U_{\rm P}(\chi) \approx \frac{1}{2}m^2\chi^2 - \frac{5}{6}\xi\frac{m^2}{M_{\rm P}^2}\chi^4 + \mathcal{O}(\chi^6) ,
\ee
from which one can see that the non-minimal coupling to gravity indeed has a flattening effect on the potential at large $\chi$.

\begin{figure}
\begin{center}
\includegraphics[width=.6\textwidth]{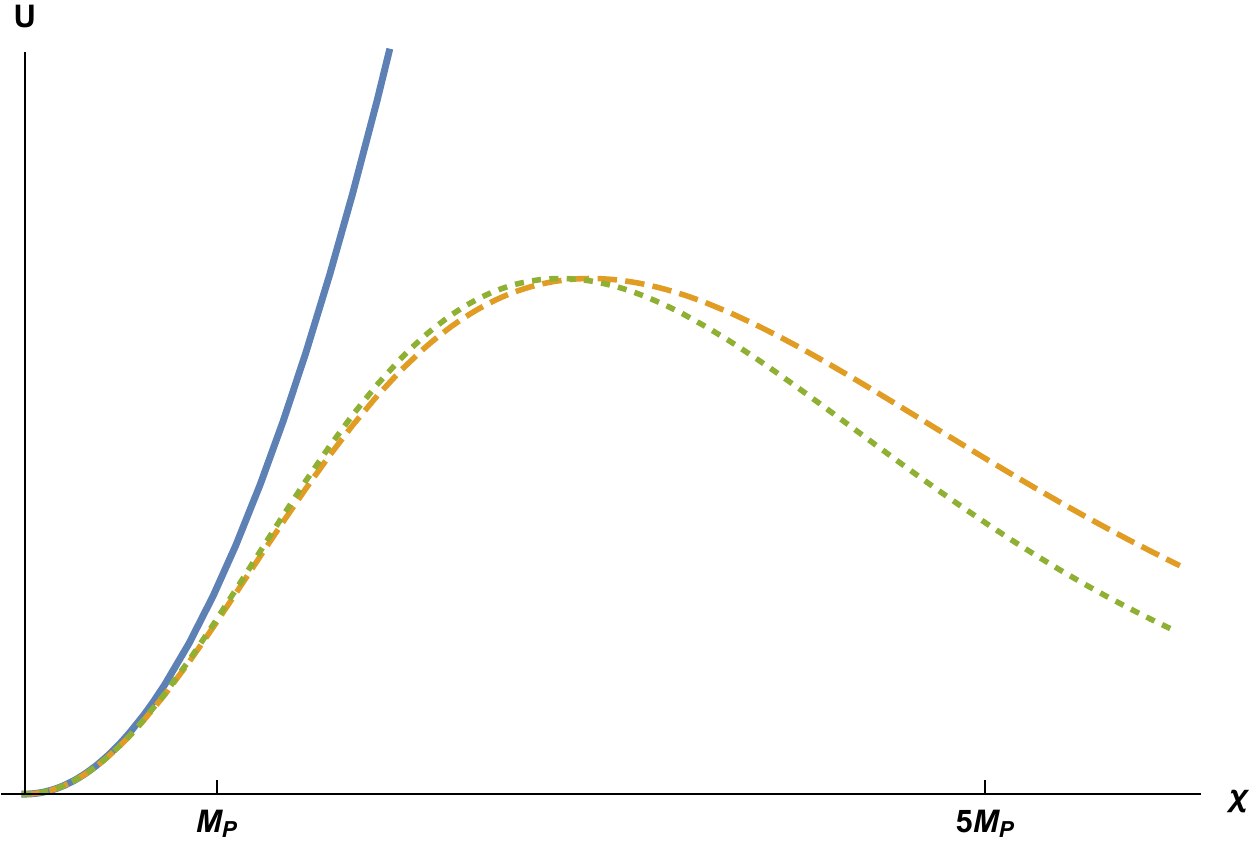}
\caption{The metric and Palatini potentials approach not only each other but also the usual quadratic case for $\xi \to 0$. The curves are the same as in Fig. \ref{potentials}. In this figure $\xi=0.1$.}
\label{potentials_smallxi}
\end{center}
\end{figure}

In the metric case the definition \eqref{chi} does not allow to determine $\phi=\phi(\chi)$ analytically but it allows expressing $\chi$ as a function of $\phi$ as \cite{GarciaBellido:2008ab}
\be
\label{chimetric}
\frac{\sqrt{\xi}\chi}{M_{\rm P}} = \sqrt{1+6\xi}{\rm sinh}^{-1}\left(\sqrt{1+6\xi}u\right) - \sqrt{6\xi}{\rm sinh}^{-1}\left(\sqrt{6\xi}\frac{u}{\sqrt{1+u^2}}\right) ,
\ee
where $u=\sqrt{\xi}\phi/M_{\rm P}$. As can be seen from the form of Eqs. \eqref{chipalatini}--\eqref{chimetric}, for $\xi\to 0$ the metric and Palatini potentials approach not only each other but also the usual quadratic case, $U=m^2\chi^2/2$, as they should (see Fig. \ref{potentials_smallxi}). However, in the following analysis we consider finite $\xi$, which allows to pinpoint the region of the model parameter space where the Quadratic Inflation can still be resurrected.

\section{Results}
\label{results}

\begin{figure}
\begin{center}
\includegraphics[width=.65\textwidth]{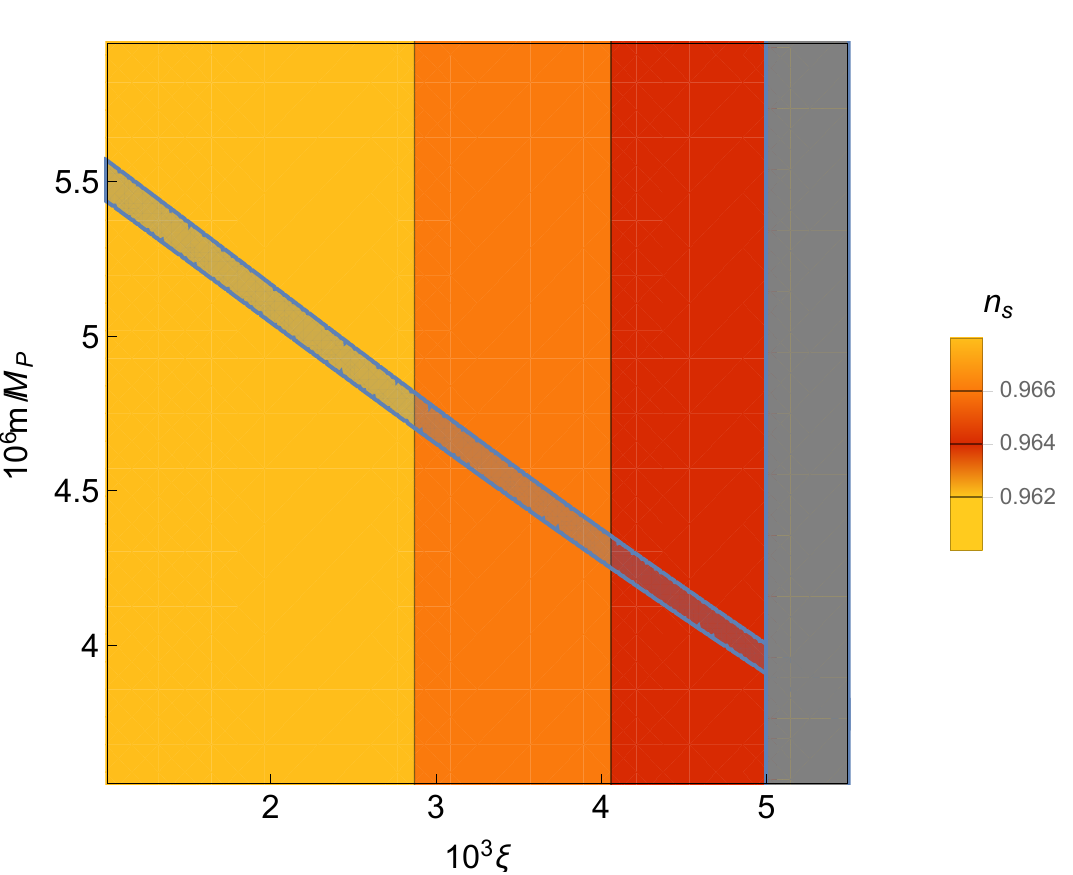}
\caption{An example of the region of the $(\xi, m)$ parameter space where the correct amplitude for the curvature power spectrum, $\mathcal{P}_{\mathcal{R}}=(2.141\pm 0.052)\times 10^{-9}$, is obtained (blue region), superimposed over contours of the spectral index $n_s$. The gray region is ruled out by $n_s<0.9681 - 0.0044$. The result is practically the same for both metric and Palatini theories of gravity. In this figure $N=60$.}
\label{PRquadr}
\end{center}
\end{figure}

\begin{figure}
\begin{center}
\includegraphics[width=.65\textwidth]{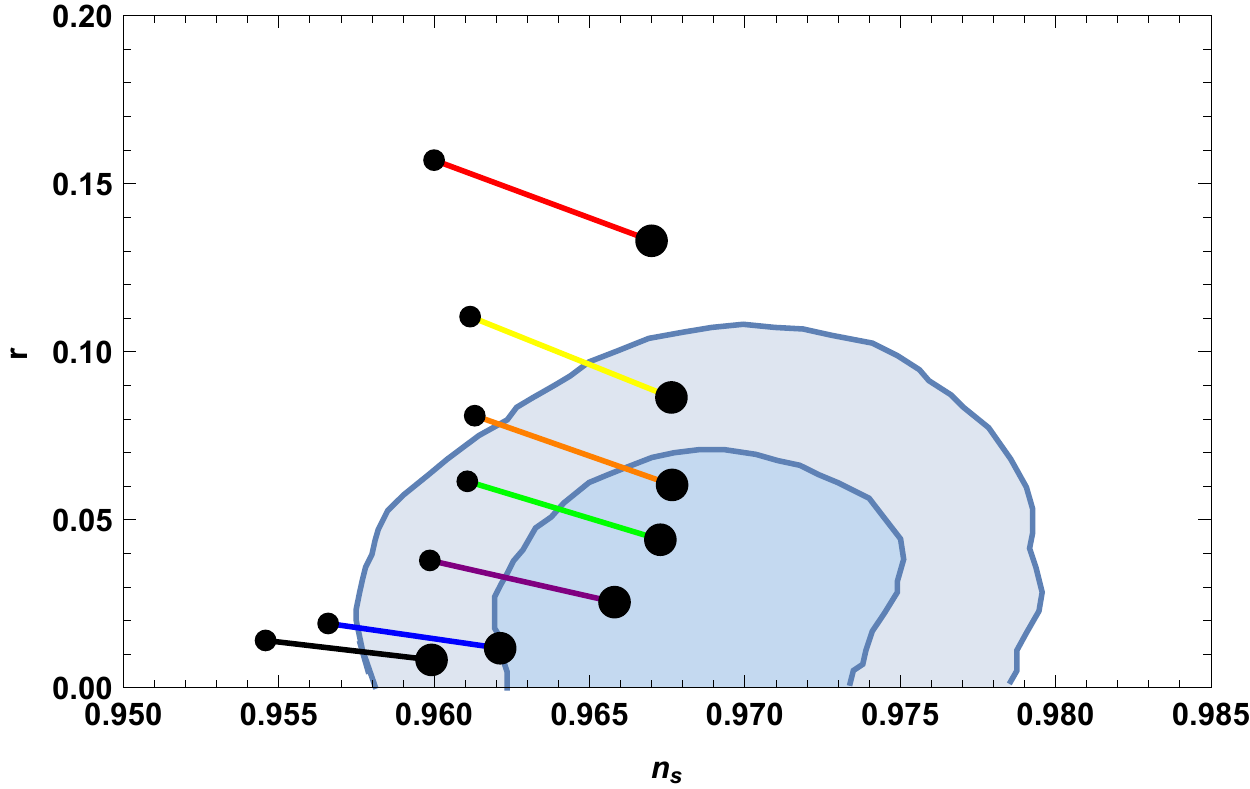}
\caption{Predictions of the models for the spectral index $n_s$ and tensor-to-scalar ratio $r$, superimposed over the $1\sigma$ and $2\sigma$ regions of the Planck TT+lowP+lensing+ext data \cite{Ade:2015xua}. The uppermost red line shows the prediction of the standard Quadratic Inflation model, whereas the yellow (second from the top), orange (third from the top), green (fourth from the top), purple (fifth from the top), blue (sixth from the top), and black (bottom) lines show the predictions of the non-minimally coupled models for $\xi=(0.6,1.2,1.8,3,5,6)\times 10^{-3}$, respectively. The results are practically the same for both metric and Palatini theories of gravity. The size of the point indicates the number of e-folds: $N=50$ for the smaller points and $N=60$ for the larger points.}
\label{nsrplot}
\end{center}
\end{figure}

Having presented how the usual Quadratic Inflation model changes in the presence of a non-minimal coupling to gravity in both the metric and Palatini cases, we now move on to study the observable consequences of such models. The minimal requirements for any successful inflation model are the following: 
\begin{enumerate}
\item The inflaton potential has to support $50-60$ e-folds of inflation.
\item The amplitude of the curvature power spectrum has to satisfy $\mathcal{P}_{\mathcal{R}}=(2.141\pm 0.052)\times 10^{-9}$ (at $1\sigma$ level), which for a given potential translates into the condition \eqref{cobe}.
\item The spectral index and tensor-to-scalar ratio have to satisfy $n_s=0.9681\pm 0.0044$ (at $1\sigma$ level) and $r<0.12$ (at $2\sigma$ level), respectively.
\end{enumerate}
We study inflation in the metric and Palatini cases by first computing the slow-roll parameters \eqref{SRparameters} for the given potentials and then, following the same procedure as presented in Sec. \ref{quadr_inf_revisited}, study for what parameter values the above conditions are met. The results are shown in the $(\xi, m)$ plane in Fig. \ref{PRquadr}, whereas Fig. \ref{nsrplot} shows the results in the usual $(n_s,r)$ plane.

\begin{figure}
\begin{center}
\includegraphics[width=.65\textwidth]{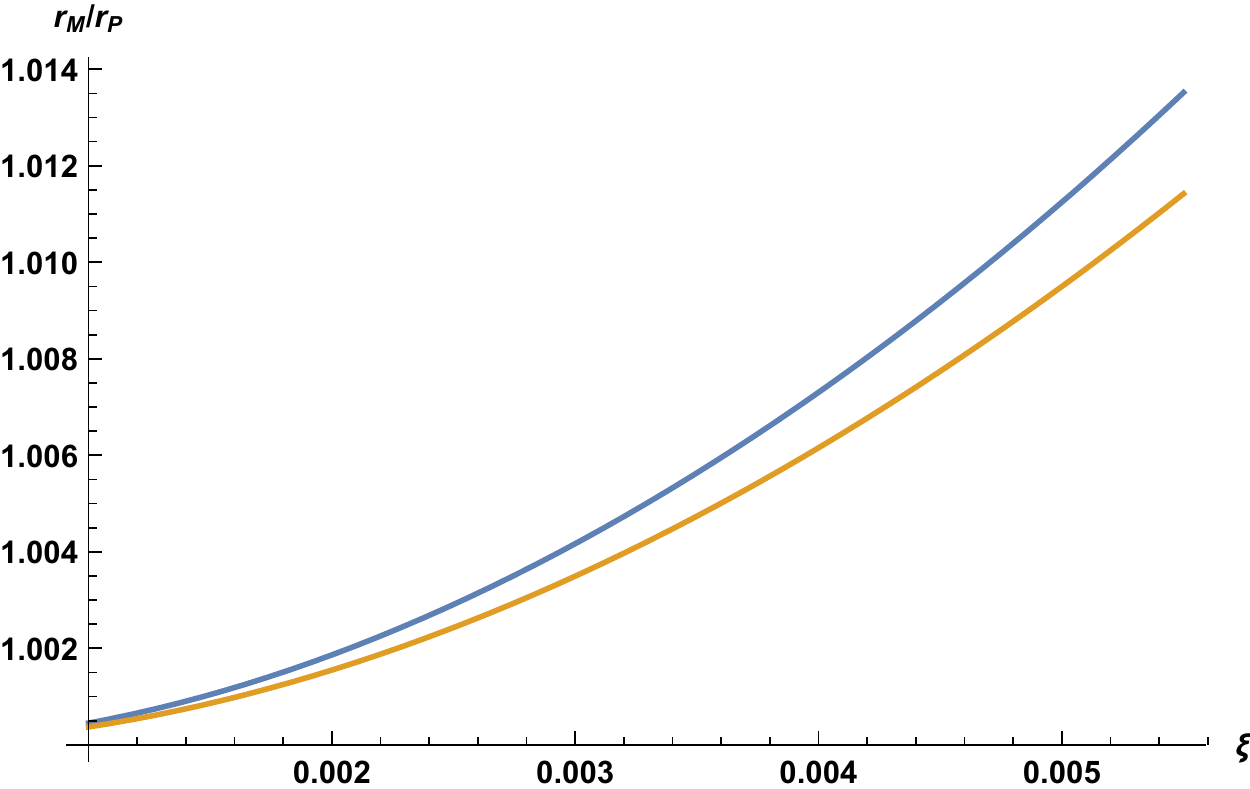}
\caption{Comparison of tensor-to-scalar ratio in the metric and Palatini cases as a function of the non-minimal coupling strength for $N=60$ (blue curve) and $N=50$ (yellow curve).}
\label{rm2rP2}
\end{center}
\end{figure}

We find in both metric and Palatini cases that only if the strength of the non-minimal gravity coupling is $1\times 10^{-3}\lesssim \xi\lesssim 5\times 10^{-3}$ for $N=60$, the models satisfy the above requirements. For $N=50$ the predictions are always off from the $1\sigma$ bound for $n_s$, see Fig. \ref{nsrplot}. For larger $\xi$ the flatness of the potential gradually decreases (see Fig. \ref{potentials}), and we find that for $\xi\gtrsim 0.4$ the potential in the metric case not only predict values for observables which are incompatible with observations but cannot support any more than $N\simeq 45$ e-folds of inflation. In the Palatini case this happens already at $\xi\simeq 0.03$.

The stringent limits by the Planck and BICEP2/Keck Array thus show that the parameter space for the non-minimal Quadratic Inflation models is very limited but non-vanishing. Interestingly, in the allowed region of the parameter space the predictions for tensor-to-scalar ratio are already within the reach of the current or near-future experiments, such as the BICEP3 \cite{Wu:2016hul} or LiteBIRD \cite{Matsumura:2013aja}. We find that in the allowed region of the parameter space the predictions is $0.01\leq r < 0.12$ for $N=60$, whereas BICEP3 and LiteBIRD aim at placing an upper bound $r\lesssim 0.03$ or $r\lesssim 0.001$, respectively, or detecting $r$ above these limits. Therefore, it is possible to either confirm the non-minimally coupled Quadratic Inflation model or rule it out in the near future.

Because for small $\xi$ the potentials in the metric and Palatini cases are very close to each other, the predictions for inflationary observables differ only very little from each other in these two cases. We find that in the region where the correct amplitude for the curvature power spectrum is obtained, the difference in the spectral index is only $\delta n_s = \mathcal{O}(10^{-5})$ and in the tensor-to-scalar ratio $\delta r = \mathcal{O}(10^{-2})$, as shown in Fig. \ref{rm2rP2}. It is therefore very challenging for the current or near-future experiments to distinguish the true gravitational degrees of freedom, at least if the action is of the form \eqref{nonminimal_action}.

However, even though in the case of a quadratic inflaton potential observations would not be able distinguish between the metric and Palatini theories of gravity, with higher-order terms in the potential, such as quartic inflaton self-interactions, $\lambda \phi^4$, this is generically not the case. In that case the Einstein frame potential exhibits a plateau at large $\chi$, and the tensor-to-scalar ratio is typically orders of magnitude larger in the metric case than in the Palatini case, as was originally studied in \cite{Bauer:2008zj} and more recently in \cite{Rasanen:2017ivk}. From this point of view, non-minimally coupled models of inflation remain as interesting probes of both high energy physics and the true nature of gravity.


\section{Conclusions}
\label{conclusions}

We studied Quadratic Inflation with a non-minimal coupling to gravity and showed that even though the standard scenario is slightly disfavored by the recent Planck data, the model can be resurrected with a suitable choice of the non-minimal coupling to gravity, $\xi=\mathcal{O}(10^{-3})$, in $V\propto m^2\phi + \xi R \phi^2$. In that case, the prediction for the tensor-to-scalar ratio is $0.01\leq r < 0.12$, which makes it possible to either confirm the scenario or rule it out already by current experiments, such as BICEP3 or LiteBIRD. We also showed that in this case the current or near-future missions are unlikely to be able to distinguish between the metric and Palatini formulations of gravity.

We showed that already a relatively small non-minimal coupling to gravity can make a large difference and make the scenario viable again. This is in contrast to the usual case of e.g. the SM Higgs inflation, where the non-minimal coupling is required to be very large, $\xi\simeq 10^4$. However, when the SM Higgs is an energetically subdominant field during inflation, small non-minimal couplings between the Higgs field and gravity, $\mathcal{O}(10^{-2})\lesssim \xi_{\rm H} \lesssim \mathcal{O}(1)$, actually become favored to avoid the Higgs from ending up to its true vacuum either during or after inflation \cite{Herranen:2014cua, Herranen:2015ima,Espinosa:2015qea,Ema:2016kpf,Kohri:2016wof,Joti:2017fwe,Figueroa:2017slm}. It is intriguing that with a non-minimal inflaton--gravity coupling not much smaller than that one can resurrect the simplest model of inflation and test the fundamental gravitational degrees of freedom.


\section*{Acknowledgements}
The author thanks P. Carrilho, M. Pieroni, H. Ramirez, and A. Salvio for correspondence and discussions and acknowledges support of the U.K. Science and Technology Facilities Council through the grant ST/J001546/1.

\bibliography{quadratic_inflation.bib}


\end{document}